\documentclass[aps,pra,twocolumn,preprintnumbers,amsmath,amssymb,floatfix,superscriptaddress]{revtex4-1}

\usepackage{graphics}
\usepackage{graphicx}
\usepackage{amsmath}
\usepackage{setspace}
\usepackage{braket}
\usepackage{epstopdf}
\usepackage{comment}
\usepackage{hyperref}
\usepackage{bm}
\usepackage{xfrac}

\hypersetup{%
   pdfpagemode=None, 
   pdfstartpage=1,
   pdfmenubar=true,
   pdftoolbar=true,
   colorlinks = true,
   linkcolor=blue,
   citecolor=blue,
   urlcolor=blue,
   bookmarksopen=false
 }

\newcommand{\affA}{Van der Waals-Zeeman Institute, Institute of Physics (IoP), University of Amsterdam, Science Park 904, 1098 XH Amsterdam, The Netherlands}

\begin{document}

\title{Spectroscopy of the $^2S_{1/2} \rightarrow\,^2P_{3/2}$ transition in Yb II: Isotope shifts, hyperfine splitting and branching ratios}

\author{T.~Feldker}\affiliation{\affA}
\author{H.~F\"urst}\affiliation{\affA}
\author{N.~V.~Ewald}\affiliation{\affA}
\author{J.~Joger}\affiliation{\affA}
\author{R.~Gerritsma}\affiliation{\affA}

\date{\today}

\begin{abstract}
We report on  spectroscopic results on the $^2S_{1/2} \rightarrow\,^2P_{3/2}$ transition in single trapped Yb$^+$ ions. We measure the isotope shifts for all stable Yb$^+$ isotopes
except $^{173}$Yb$^+$, as well as the hyperfine splitting of the $^2P_{3/2}$ state in $^{171}$Yb$^+$. Our results are in agreement with previous measurements but are a factor of 5-9 more precise. For the hyperfine constant $A\left(^2P_{3/2}\right) = 875.4(10)$\,MHz our results also agree with previous measurements but deviate significantly from theoretical predictions.
We present experimental results on the branching ratios for the decay of the $^2P_{3/2}$ state. We find branching fractions for the decay to the $^2D_{3/2}$ state and $^2D_{5/2}$ state of 0.17(1)\% and 1.08(5)\%, respectively, in rough agreement with theoretical predictions. 
Furthermore, we measured the isotope shifts of the $^2F_{7/2} \rightarrow\,^1D\left[5/2\right]_{5/2}$ transition and determine the hyperfine structure constant for the $^1D\left[5/2\right]_{5/2}$ state in $^{171}$Yb$^+$ to be $A(^1D\left[5/2\right]_{5/2}) = -107(6)$\,MHz.
\end{abstract}


\maketitle

\section{Introduction}

Laser-cooled ions in Paul traps form one of the most mature laboratory systems for performing optical metrology, precision measurements as well as quantum-computation and quantum-simulation\,\cite{Diddams:2001,Fortson:1993,Haffner:2000,Cirac:1995a,Blatt:2012}. The ion species Yb$^+$ is a particularly versatile system for many of these applications owing to its rich electronic structure with multiple meta-stable states\,\cite{Tamm:2009,Godun:2014}. Furthermore, the hyperfine structure of $^{171}$Yb$^+$ provides a first-order magnetic field-insensitive qubit in the electronic ground state\,\cite{Blatt:1983, Fisk:1997, Olmschenk:2007} that may be used in quantum information applications~\cite{Islam:2013,Debnath:2016}.

While many transitions between low-lying electronic states in Yb$^+$ have been studied with great precision\,\cite{Taylor:1997, Taylor:1999, Yu:2000, Olmschenk:2007, Tamm:2007}, there has been only one measurement of the isotope shifts in the $^2S_{1/2} \rightarrow\,^2P_{3/2}$ (D2) transition as well as of the hyperfine splitting of the $^2P_{3/2}$ state\,\cite{Berends:1992} so far, which was performed in a hollow-cathode discharge lamp. Remarkably, the experimental result for the hyperfine splitting disagrees with theoretical predictions\,\cite{Pendrill:1994, Safronova:2009, Porsev:2012, Sahoo:2011, Dzuba:2011}. Although there has been a lot of theoretical work on transition amplitudes for the decay of the $^2P_{3/2}$ state\,\cite{Safronova:2009, Biemont:1998, Porsev:2012, Sahoo:2011, Roberts:2014}, there seems to be no experimental data available for the branching ratios of the decay of the $^2P_{3/2}$ state up until now.

Here, we present experimental results on the isotope shifts in the D2-transition, the hyperfine splitting of the $^2P_{3/2}$ state as well as on the branching ratios of its decay obtained from a single trapped and laser-cooled ion. Furthermore, we present measurements of the isotope shift in the $^2F_{7/2} \rightarrow\,^1D\left[5/2\right]_{5/2}$ transition, as well as the hyperfine splitting in the $^1D\left[5/2\right]_{5/2}$ state in $^{171}$Yb$^+$. Single trapped ions are very well suited to perform such precision measurements, as both state preparation and detection can be performed with great accuracy while at the same time errors due to back-ground gas collisions are negligible.  Using isotope-selective photo-ionization to load the Paul trap, we are able to conduct the experiments even with the rare isotope $^{168}$Yb$^+$ (0.13\% abundance,\cite{NIST:Yb}) for which no previous data seems to exist.

\section{Experimental setup}

The experiments have been performed in a linear Paul trap described in Ref.\,\cite{Joger:2017}. We load a single Yb$^+$ ion into the trap by two-step photo-ionization with lasers at 399\,nm wavelength for the resonant excitation of the $^1S_0 \rightarrow\,^1P_1$ transition in neutral Yb and 369\,nm wavelength for the excitation into the ionization continuum. Tuning the wavelength of the first step to the resonance of a specific isotope allows for isotope selective loading of Yb$^+$ ions. Due to overlapping resonances\,\cite{Kleinert:2016}, $^{170}$Yb$^+$ and $^{172}$Yb$^+$ cannot be loaded deterministically, but only in combination with $^{171}$Yb$^+$ and $^{173}$Yb$^+$, respectively. However, by temporarily lowering the trap drive amplitude we can expel the heavier isotopes from the trap and keep only the isotope $^{170}$Yb$^+$ or $^{172}$Yb$^+$, respectively.

 \begin{figure}[htbp]
	\centering
	\includegraphics[width=0.95\columnwidth]{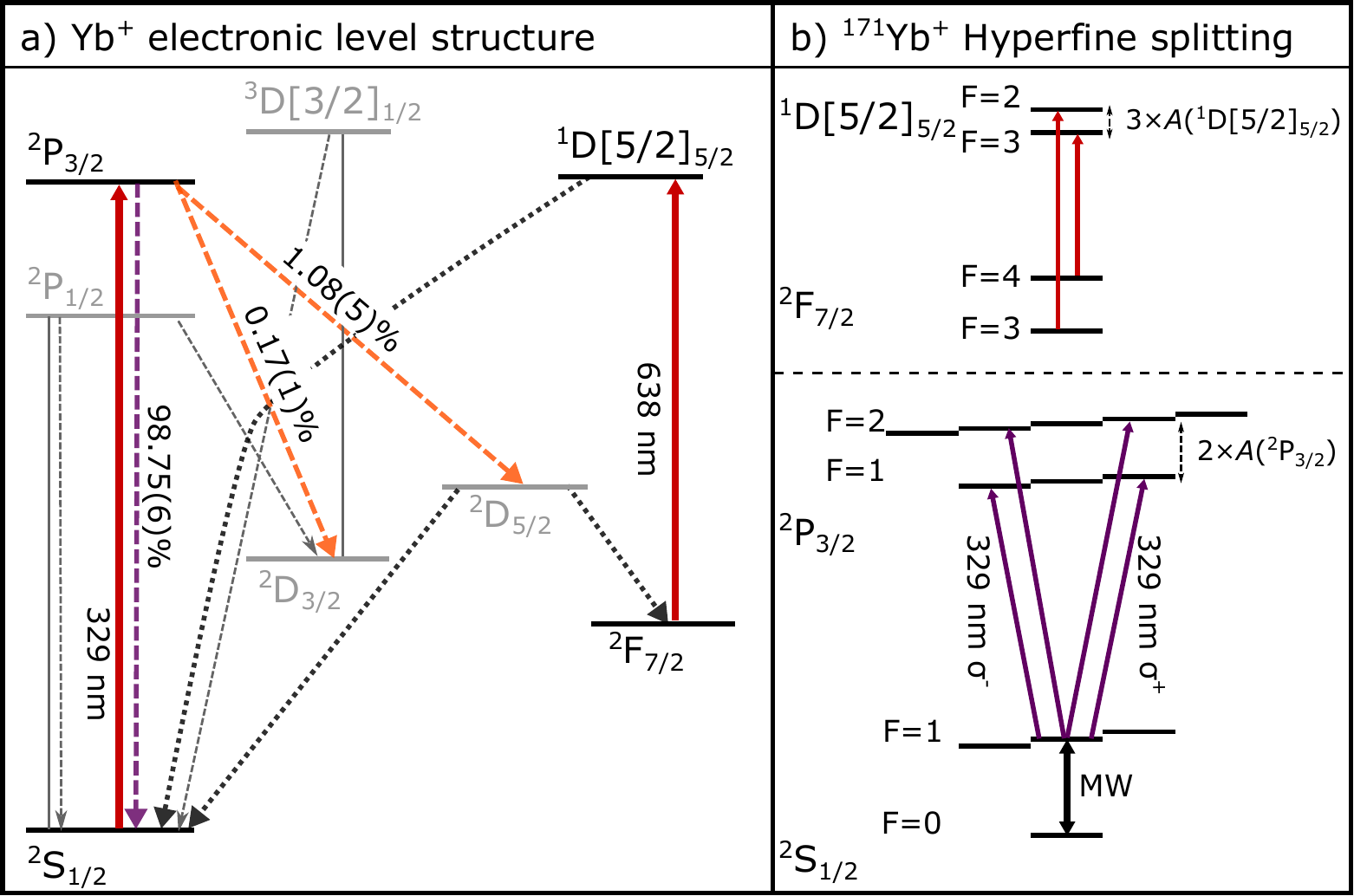}
	\caption{a) Relevant electronic levels and transitions in Yb$^+$. We perform spectroscopic measurements on the $^2S_{1/2} \rightarrow\,^2P_{3/2}$ transition near 329\,nm wavelength. The $^2P_{3/2}$ state decays in $\tau = 6.15(9)$\,ns\,\cite{Pinnington:1997} either back into the ground state or into one of the metastable states $^2D_{3/2}$ or $^2D_{5/2}$. While an ion which is initially in the states $^2S_{1/2}$ or $^2D_{3/2}$ scatters light during Doppler cooling (thin gray arrows), it will not scatter light when it is in the $^2D_{5/2}$ state. This allows for detection of a successful excitation of the $^2S_{1/2} \rightarrow\,^2P_{3/2}$ transition as well as for measurement of the branching fractions of the $^2P_{3/2}$ decay. From the $^2D_{5/2}$ state the ion decays in $\tau = 7.2(3)$\,ms\,\cite{Taylor:1997} to either the ground state or the very long lived $^2F_{7/2}$ state ($\tau \approx 10$\,yr\,\cite{Roberts:1997}). We use the $^2F_{7/2} \rightarrow\,^1D\left[5/2\right]_{5/2}$ transition near 638\,nm wavelength to depopulate the $^2F_{7/2}$ state. b) Hyperfine structure of the $^2F_{7/2} \rightarrow\,^1D\left[5/2\right]_{5/2}$ and $^2S_{1/2} \rightarrow\,^2P_{3/2}$ transitions in $^{171}$Yb$^+$. }
	\label{fig:Scheme}
\end{figure}

 Lasers near wavelengths of 369\,nm and 935\,nm are used to Doppler cool the ion on the $^2S_{1/2} \rightarrow\,^2P_{1/2}$ transition and pump population trapped in the metastable $^2D_{3/2}$ state back into the cooling cycle via excitation to the $^3D\left[3/2\right]_{1/2}$ state, as shown in Fig.\,\ref{fig:Scheme}a. We image the ion's fluorescence at 369\,nm wavelength to a photo-multiplier-tube (PMT) for detection.

 For cooling of the isotope $^{171}$Yb$^+$, which has a nuclear spin of $I=1/2$ and accordingly hyperfine splittings of the electronic states, we use the closed transition $\left|^2S_{1/2},F=1\right\rangle\rightarrow\left|^2P_{1/2},F=0\right\rangle$. However, due to off-resonant excitation of the $\left|^2P_{1/2},F=1\right\rangle$ state, the ion occasionally decays to the $\left|^2S_{1/2},F=0\right\rangle$ state. Microwave radiation at 12.6\,GHz couples the $F=0$ and $F=1$ ground states to ensure continuous cooling. In order to prepare the ion in the $\left|^2S_{1/2},F=0\right\rangle$ state, we excite the $\left|^2S_{1/2},F=1\right\rangle \rightarrow \left|^2P_{1/2},F=1\right\rangle$ transition resonantly while the microwave radiation is switched off.

 \begin{figure}[htbp]
 	\centering
 	\includegraphics[width=0.75\columnwidth]{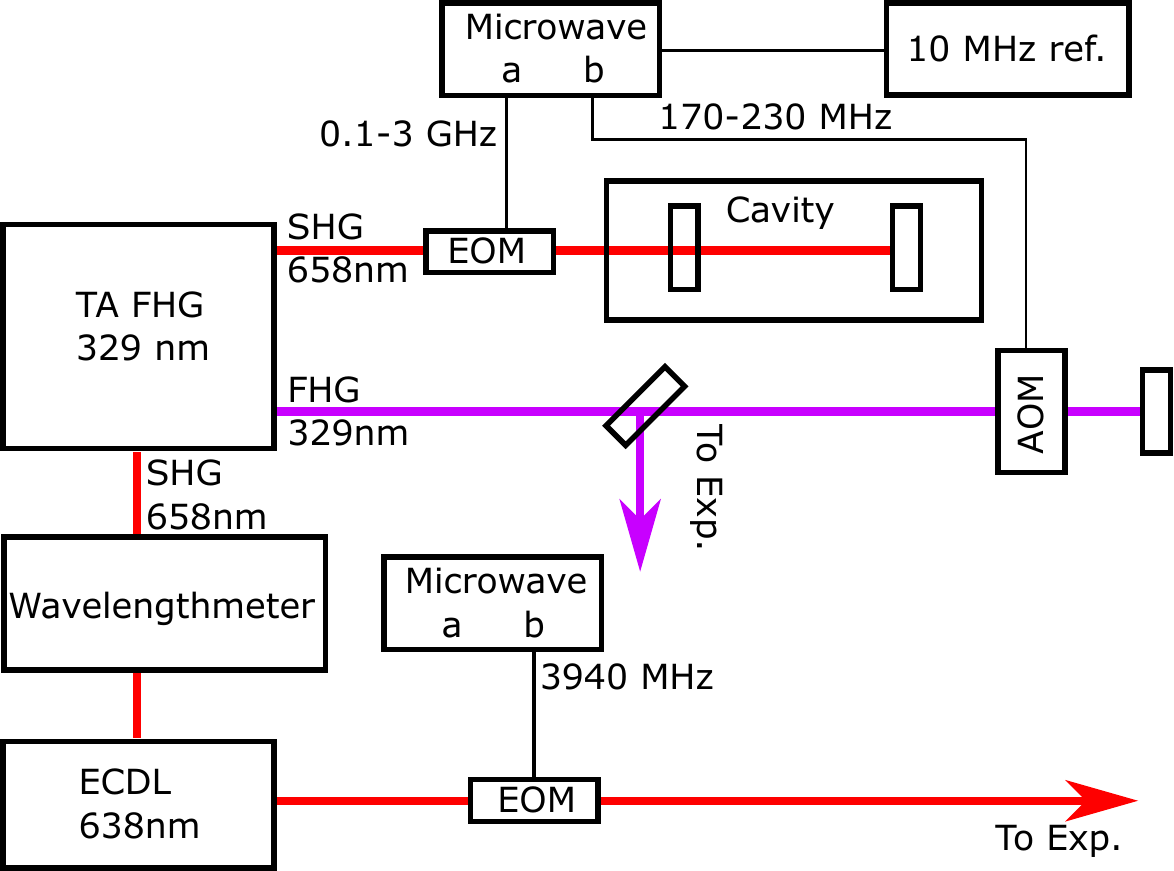}
 	\caption{Sketch of the laser setup for excitation of the $^2F_{7/2} \rightarrow\,^1D\left[5/2\right]_{5/2}$ and $^2S_{1/2} \rightarrow\,^2P_{3/2}$ transitions. Radiation near 329\,nm wavelength is generated by a frequency quadrupled, amplified diode laser. After the first frequency doubling, a part of the light is coupled out and used for frequency stabilization to a fixed reference cavity. Using a fiber coupled electro-optical-modulator (EOM), we generate sidebands with a frequency between 0.1-3\,GHz which are used to offset the laser's carrier frequency with respect to the cavity mode. We use an acousto-optical-modulator (AOM) in double pass configuration for pulse shaping and frequency scanning.
 	The laser at 638\,nm wavelength is a home-made external cavity diode laser (ECDL) with a grating in Littrow configuration. It is frequency stabilized to the wavelength meter and switched by a mechanical shutter. Using a fiber EOM, sidebands can be modulated in order to drive transitions between multiple hyperfine states.}
 	\label{fig:setup}
 \end{figure}

 \begin{table*}[tbp]
 	\centering
 	\begin{tabular}{cccc}
 		\multicolumn{1}{c|}{\textbf{Yb$^+$ Isotope}} & \multicolumn{1}{c|}{\textbf{$^2P_{3/2}$ -- this work}}&\multicolumn{1}{c|}{\textbf{$^2P_{3/2}$ -- Ref.\,\cite{Berends:1992}}}& \multicolumn{1}{c|}{\textbf{ $^2F_{7/2}\rightarrow\,^1D\left[5/2\right]_{5/2}$ -- this work}} \\[0.3ex]
 		\hline
 		\multicolumn{1}{c|}{\textbf{168}} & \multicolumn{1}{c|}{3007.8(30)\,MHz} & \multicolumn{1}{c|}{--}& \multicolumn{1}{c|} {6.04(2)\,GHz} \\  \hline
 		\multicolumn{1}{c|}{\textbf{170}} & \multicolumn{1}{c|}{1457.9(30)\,MHz}& \multicolumn{1}{c|}{1459(21)\,MHz}&  \multicolumn{1}{c|}{2.93(2)\,GHz} \\  \hline
 		\multicolumn{1}{c|}{\textbf{171}} & \multicolumn{1}{c|}{922.5(25)\,MHz}& \multicolumn{1}{c|}{920(15)\,MHz}&  \multicolumn{1}{c|}{1.87(2)\,GHz} \\  \hline
 		\multicolumn{1}{c|}{\textbf{172}} & \multicolumn{1}{c|}{0}& \multicolumn{1}{c|}{0}&  \multicolumn{1}{c|}{0} \\  \hline	
 		\multicolumn{1}{c|}{\textbf{174}} & \multicolumn{1}{c|}{-1152.3(15)\,MHz}& \multicolumn{1}{c|}{-1154(11)\,MHz}&  \multicolumn{1}{c|}{-2.26(2)\,GHz} \\  \hline
 		\multicolumn{1}{c|}{\textbf{176}} & \multicolumn{1}{c|}{-2254.8(15)\,MHz}& \multicolumn{1}{c|}{-2259(13)\,MHz}&  \multicolumn{1}{c|}{-4.41(2)\,GHz} \\  \hline	
 	\end{tabular}
 	\caption{Isotope shifts of the $^2P_{3/2}$ state in Yb$^+$ as measured by our group compared to values from Ref.\,\cite{Berends:1992}. The number in brackets denotes the error in the last digit. The isotope shift of the $^2F_{7/2}\rightarrow\,^1D\left[5/2\right]_{5/2}$ transition as measured with the wavelength meter is given in the last column.}
 	\label{tab:IsotopeShift}
 \end{table*}

 In addition to the lasers required for cooling and detection of the ion, we use light near the wavelengths of 329\,nm and 638\,nm to drive the transitions $^2S_{1/2} \rightarrow\,^2P_{3/2}$ and $^2F_{7/2} \rightarrow\,^1D\left[5/2\right]_{5/2}$, respectively.
 We generate light at 329\,nm wavelength with a frequency quadrupled, amplified diode laser. After the first doubling cavity, light at 658\,nm wavelength is coupled into a high bandwidth fiber-electro-optical modulator (EOM). Sidebands at frequencies of 0.1-3\,GHz are modulated onto the light and used to stabilize the laser to an external reference cavity. Thus, the laser is stabilized to the fixed reference cavity with a variable frequency offset which is given by the modulation frequency of the EOM. The reference cavity consists of two mirrors with a reflectivity of $R\approx 99\%$ glued to a 10\,cm long Zerodur spacer in a temperature stabilized vacuum housing. The laser is frequency stabilized by the Pound-Drever-Hall technique~\cite{Drever:1983}.

 For further frequency scanning and pulse shaping we use an acousto-optical modulator (AOM) in double-pass configuration with a center frequency of 200\,MHz and a bandwidth of 100\,MHz. The signals for the AOM and the fiber EOM are generated by a two channel microwave generator, which is stabilized to a 10\,MHz reference signal from a signal generator.
 A mechanical shutter prevents any light from reaching the ion if switched off. Light from the first doubling cavity is coupled to a commercial wavelength meter, allowing for a course absolute frequency determination of the frequency-quadrupled light with an accuracy of 60\,MHz according to specification.

 We generate light at 638\,nm wavelength with a home-built ECDL. This laser is stabilized to the wavelength meter to compensate for frequency drifts and has a short-time frequency stability of better than 10\,MHz. We switch the light with a mechanical shutter. The light is coupled to a fiber EOM which allows for modulating sidebands in order to drive transitions between multiple hyperfine states in the case of $^{171}$Yb$^+$, and guided to the experiment.
 The part of the setup relevant for the spectroscopy is shown in Fig.\ref{fig:setup}.

\section{Results}

\subsection{Isotope shifts and hyperfine splitting}

\begin{figure}[htbp]
	\centering
	\includegraphics[width=0.95\columnwidth]{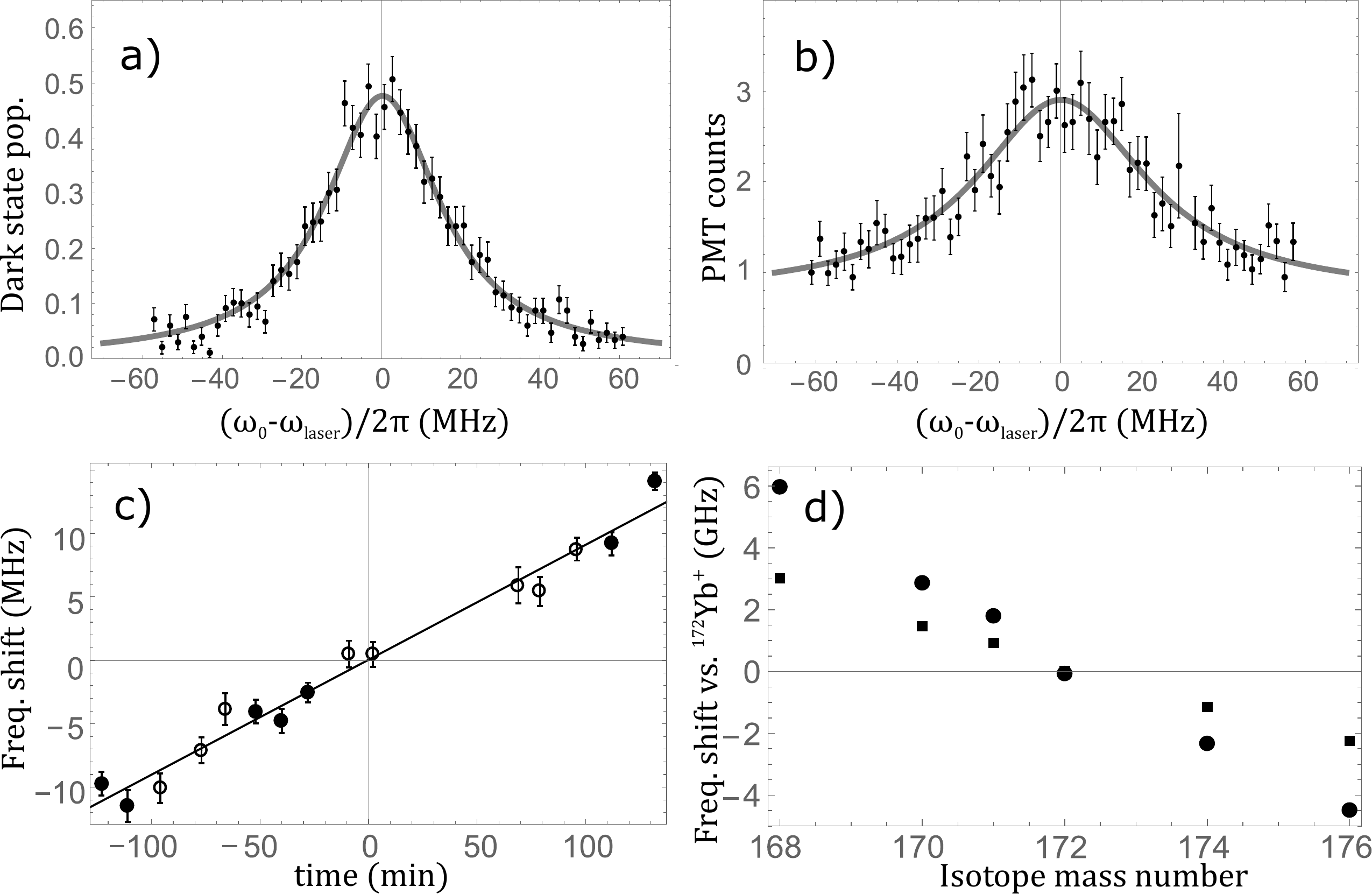}
	\caption{Isotope shift measurement with trapped Yb$^+$ ions.  \textbf{a)} Data from a single scan over the $^2S_{1/2}\rightarrow\,^2P_{3/2}$ transition in $^{174}$Yb$^+$. Error bars denote the quantum projection noise. \textbf{b)} Plot of a single resonance scan in $^{171}$Yb$^+$. Here, we do not project to bright or dark states but measure PMT counts during the detection bin ($t_{\mathrm{det}} = 2.5$\,ms). The error bars denote the uncertainties from photon counting statistics. \textbf{c)} Cavity drift obtained by repeated measurements of the transitions from $\left|^2S_{1/2},F=1\right\rangle$ to $\left|^2P_{3/2},F=1\right\rangle$ (circles) or $\left|^2P_{3/2},F=2\right\rangle$ (disks). Offsets of $\nu_1=1811.0$\,MHz ($F=1$, circles) and $\nu_2=-60.0$\,MHz ($F=2$, disks) are added to the measured resonance frequencies. We choose $\nu_1$ and $\nu_2$ so that the standard deviation for a combined linear fit (black line) is minimized. From the gradient of the fit, we determine a linear drift of the cavity of 92\,kHz/min. The hyperfine splitting of the $^2P_{3/2}$ state is given by the difference $\nu_1-\nu_2=1751.0\,\mathrm{MHz}$. The result given in Tab.\,\ref{tab:HFSsplitting} is based on the average of two measurement series on different days. \textbf{d)} Isotope shifts in the $^2S_{1/2} \rightarrow\,^2P_{3/2}$ transition (circles) and the $^2F_{7/2} \rightarrow\,^1D\left[5/2\right]_{5/2}$ transition (squares). Error bars are too small to be visible on this scale.}
	\label{fig:isoShift}
\end{figure}

 In case of the isotopes without nuclear spin, we measure the resonance frequency of the D2-transition by applying laser pulses with a width of 5\,$\mu$s and a saturation parameter $s \approx 1$ to the ion.
 From the $^2P_{3/2}$ state, there is a probability of about 1\% for the ion to decay to the metastable state $^2D_{5/2}$ ($\tau=7.2$\,ms) from where it decays with 83\% probability\,\cite{Taylor:1997} to the long-lived $^2F_{7/2}$ state.
 An ion in either of these states does not scatter light during Doppler cooling.
 On resonance, about 50\% of the population decays to the dark $^2D_{5/2}$ state in 5\,$\mu$s.
 To detect whether the ion is in one of the dark states, we image the ion's fluorescence to a PMT for 4\,ms during Doppler cooling, allowing for almost perfect state detection. We scan the laser over the atomic resonance in steps of 2\,MHz by tuning the drive frequency of the AOM ($\nu_{\mathrm{aom}}$).
 We compensate for the frequency-dependence of the diffraction efficiency in the AOM by supplying appropriate radio-frequency power at each frequency.
 After the detection, we pump the ion back into the cooling cycle by exciting the $^2F_{7/2} \rightarrow\,^1D\left[5/2\right]_{5/2}$ transition.
 A post-selection measurement is performed before each spectroscopy pulse in order to check if the ion was successfully pumped out of the $^2F_{7/2}$ state. The measurement data for a single scan of the transition in $^{174}$Yb$^+$ is plotted in Fig.\,\ref{fig:isoShift}a.

We repeat the experiment for the isotopes $^{168}$Yb$^+$, $^{170}$Yb$^+$, $^{172}$Yb$^+$, $^{174}$Yb$^+$ and $^{176}$Yb$^+$. For each measurement, we frequency-stabilize the laser to the same cavity mode, but with different offset frequencies $\nu_{\mathrm{eom}}$ given by the modulation frequency of the fiber EOM. The relative frequency of the spectroscopy light compared to the fixed cavity resonance is determined by $\nu_{\mathrm{rel}} = 2 \nu_{\mathrm{eom}} - 2 \nu_{\mathrm{aom}}$. We estimate the drift of the cavity by measuring the same resonance at different times as shown in Fig.\,\ref{fig:isoShift}c.

The uncertainties in the energy shifts are given by the statistical error of the fitted resonance and the uncertainty in the cavity drift during the measurement time. In principle the error should not depend on the abundance of the isotope. However, the measurements with the rare isotopes $^{168}$Yb$^+$ (0.13\% abundance) and $^{170}$Yb$^+$ (3.05\% abundance, can only be loaded in combination with $^{172}$Yb$^+$)\,\cite{NIST:Yb} take significantly longer due to low ion loading rates. This leads to larger uncertainties for the cavity drift. We use $\pi$-polarized light and a small magnetic field of $B = 0.05$\,mT to avoid errors due to Zeeman-shifts of the D2-transition. The results are shown in Fig.\,\ref{fig:isoShift}d and Tab.\,\ref{tab:IsotopeShift}.

\begin{table*}[tbp]
	\centering
	\begin{tabular}{ccccccccc}
		\multicolumn{1}{l|}{}  & \multicolumn{1}{c|}{\textbf{\cite{Pendrill:1994}}$_{\text{Th.}}$} &\multicolumn{1}{l|}{\textbf{\cite{Safronova:2009}}$_{\text{Th.}}$}&\multicolumn{1}{l|}{\textbf{\cite{Porsev:2012}$^a$}$_{\text{Th.}}$}&\multicolumn{1}{l|}{\textbf{\cite{Porsev:2012}$^b$}$_{\text{Th.}}$}&\multicolumn{1}{l|}{\textbf{\cite{Sahoo:2011}}$_{\text{Th.}}$}&\multicolumn{1}{l|}{\textbf{\cite{Dzuba:2011}}$_{\text{Th.}}$} &\multicolumn{1}{l|}{\textbf{\cite{Berends:1992}}$_{\text{Exp.}}$}&\multicolumn{1}{l|}{\textbf{This work}}\\
		\hline
		\multicolumn{1}{l|}{\textbf{$A\left(^2P_{3/2}\right)$}} & \multicolumn{1}{c|}{391} & \multicolumn{1}{c|}{311.5}&\multicolumn{1}{c|}{330}&\multicolumn{1}{c|}{765}&\multicolumn{1}{c|}{322}&\multicolumn{1}{c|}{388}&\multicolumn{1}{l|}{877(16)}& \multicolumn{1}{c|}{\textbf{875.4(10)}}\\  \hline
		\multicolumn{1}{l|}{\textbf{$A(^1D\left[5/2\right]_{5/2})$}}&\multicolumn{1}{c|}{-}&\multicolumn{1}{c|}{-}&\multicolumn{1}{c|}{-}&\multicolumn{1}{c|}{-}&\multicolumn{1}{c|}{-} & \multicolumn{1}{c|}{-}& \multicolumn{1}{c|}{-}& \multicolumn{1}{c|}{\textbf{-107(6)}} \\  \hline	
	\end{tabular}

\caption{Hyperfine structure constants $A$ in $^{171}$Yb$^+$ in~MHz. The number in brackets denotes the error in the last digit. The magnetic-dipole hyperfine structure constant $A\left(^2P_{3/2}\right)$ measured by our group is consistent with previous measurements (Exp.) but deviates significantly from theory (Th.) predictions. In Ref.\,\cite{Porsev:2012} results are given for a single-electron approach ($a$) and a many-electron approach ($b$). The many-electron results give the best agreement with our experimental results by far.}
	\label{tab:HFSsplitting}
\end{table*}

In case of $^{171}$Yb$^+$ we make use of the hyperfine structure (see Fig.\,\ref{fig:Scheme}b) in order to detect the excitation to the $^2P_{3/2}$ state. We prepare the ion in the $\left|F=0\right\rangle$ ground state before transferring it to the $\left|F=1, m_F=0\right\rangle$ state via rapid adiabatic passage (RAP) using microwave radiation. We apply a laser pulse with a width of 200\,ns to excite the $\left|^2P_{3/2},F=1\right\rangle$ or $\left|^2P_{3/2},F=2\right\rangle$ states, followed by a second RAP pulse on the $\left|F=0\right\rangle\rightarrow\left|F=1,m_F=0\right\rangle$ transition. At the end of this sequence, we perform state-selective fluorescence detection, based on Doppler cooling without microwave coupling of the $\left|F=1\right\rangle$ and $\left|F=0\right\rangle$ ground states. During the detection, an ion in the $\left|F=0\right\rangle$ state appears dark while an ion in the $\left|F=1\right\rangle$ state scatters light, which allows for detection of the induced population transfer out of the initial $\left|F=1, m_F=0\right\rangle$ state.

Experiments with $^{171}$Yb$^+$ are conducted in a magnetic field of 0.18-0.23\,mT in order to allow for efficient Doppler cooling on the $\left|^2S_{1/2},F=1\right\rangle \rightarrow \left|^2P_{1/2},F=0\right\rangle$ transition\,\cite{Olmschenk:2007}.
We use linear polarized light ($\sigma^++\sigma^-$) at 329\,nm wavelength to avoid the dipole forbidden $\left|^2S_{1/2},F=1,m_F=0\right\rangle \rightarrow \left|^2P_{3/2},F=1,m_F=0\right\rangle$ transition. Due to the symmetric excitation of the transition (see Fig.\,\ref{fig:Scheme}b) the magnetic field should not lead to a frequency shift of the transition. Measurements at different fields corroborate this assumption.

We repeat the experiment for both hyperfine states and determine the hyperfine energy splitting given in Tab.\ref{tab:HFSsplitting}. Together with the well-known energy splitting of the ground state of 12642.812\,MHz\,\cite{Fisk:1997} we calculate the isotope shift of the $^2P_{3/2}$ state given in Tab.\,\ref{tab:IsotopeShift}.

During the spectroscopy of the $^2S_{1/2} \rightarrow\,^2P_{3/2}$ transition, the $^2F_{7/2}$ state with a lifetime of $\tau > 10\, $yr\,\cite{Roberts:1997} is populated via decay of the $^2D_{5/2}$ state.
After the state-detection, we pump population out of the $^2F_{7/2}$ state by excitation of the $^2F_{7/2} \rightarrow\,^1D\left[5/2\right]_{5/2}$ transition. Wavelength meter readings during the measurement give us the isotope shifts of this transition with an accuracy of 20\,MHz.
To the best of our knowledge, this is the most complete (in terms of measured isotopes) and precise measurement of the isotope shifts of the $^2F_{7/2} \rightarrow\,^1D\left[5/2\right]_{5/2}$ transition. In $^{171}\mathrm{Yb}^+$ we drive the transitions $\left|F=4\right\rangle \rightarrow \left|F=3\right\rangle$ and $\left|F=3\right\rangle \rightarrow \left|F=2\right\rangle$ (see Fig.\,\ref{fig:Scheme}b).
We use sidebands at 3940\,MHz to excite both transitions efficiently. With a hyperfine splitting of the $^2F_{7/2}$ state of 3620(2)\,MHz\,\cite{Taylor:1999}, we determine an energy splitting of the upper $^1D\left[5/2\right]_{5/2}$  state of 320(20)\,MHz and a hyperfine constant $A(^1D\left[5/2\right]_{5/2})= -107(6)$\,MHz.
There seems to be no previous experimental data available for the $^1D\left[5/2\right]_{5/2}$ hyperfine splitting, only a theoretical estimation of $A= 199$\,MHz\,\cite{Petrasiunas:2011}, which deviates significantly from the value we find in the experiment.

\begin{figure}[htbp]
	\centering
	\includegraphics[width=0.95\columnwidth]{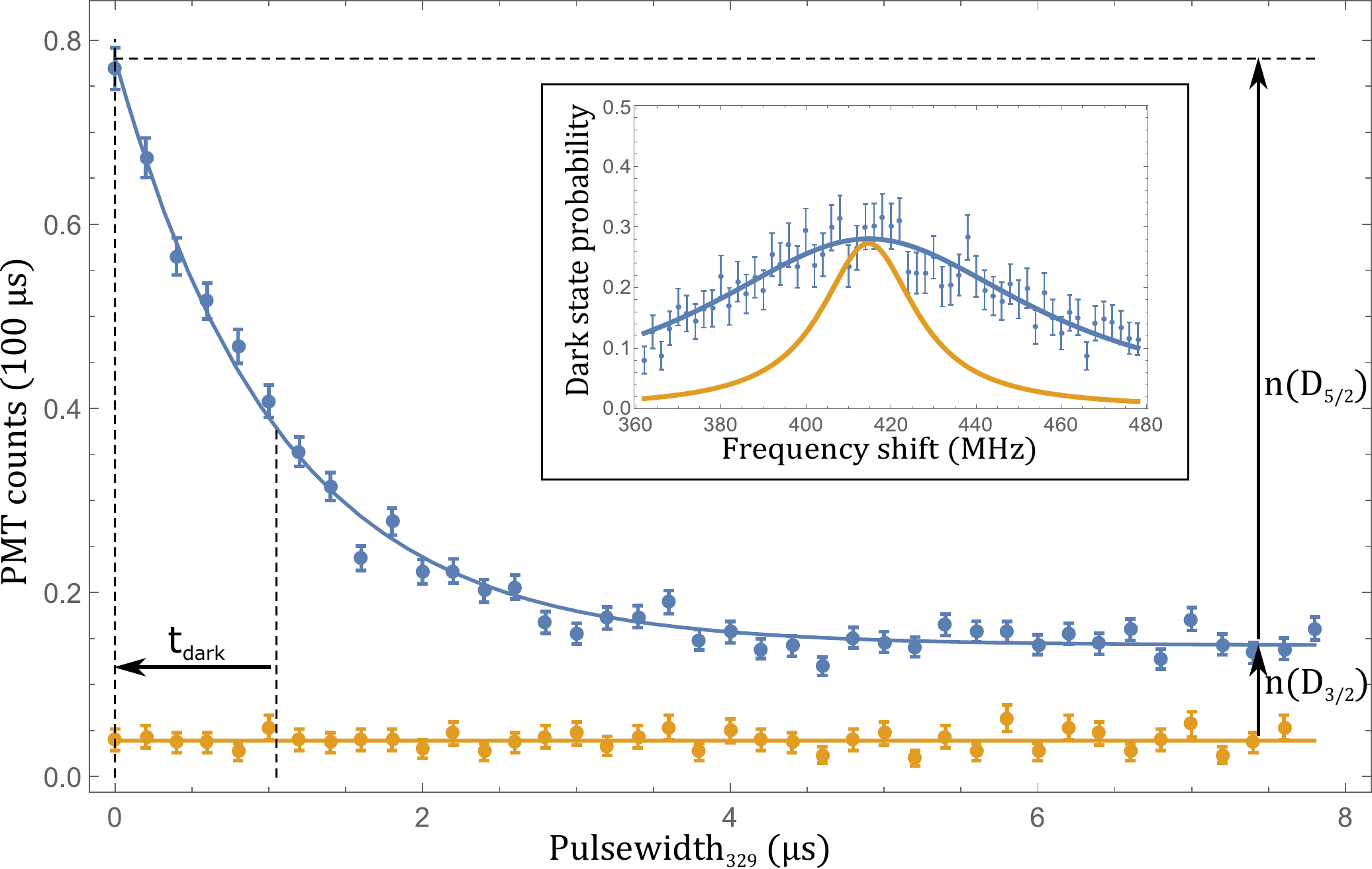}
	\caption{Decay measurement of the $P_{3/2}$ state. Plot of PMT counts during a 100\,$\mu$s detection bin versus pulse width of a laser pulse, resonant with the D2-transition (blue) and background counts measured without an ion in the trap during the same 100\,$\mu$s detection bin (yellow). From the decay time constant $\tau_{\mathrm{dark}}$ we determine the combined decay probability to the $D$ states. From the ratio of fluorescence levels $n({\mathrm{D_{5/2}}})/n({\mathrm{D_{3/2}}})$, we determine the relative strength of the decays to the $D$ manifold. The inset shows a scan over the $^2S_{1/2}\rightarrow\,^2P_{3/2}$ resonance at a saturation parameter of $s=11$ (blue). For comparison the calculated non-saturated line is shown (yellow).}
	\label{fig:Branching}
\end{figure}

\subsection{Branching fractions}

\begin{table*}[tbp]
	\centering
	\begin{tabular}{ccccccc}
		\multicolumn{1}{c|}{\textbf{Branching from $P_{3/2}$ to:}} & \multicolumn{1}{c|}{\cite{Biemont:1998}$_{\text{Th.}}$}&\multicolumn{1}{c|}{\cite{Porsev:2012}$^a$$_{\text{Th.}}$}&\multicolumn{1}{c|}{\cite{Porsev:2012}$^b$$_{\text{Th.}}$}&\multicolumn{1}{c|}{\cite{Sahoo:2011}$_{\text{Th.}}$}&\multicolumn{1}{c|}{\cite{Roberts:2014}$_{\text{Th.}}$}& \multicolumn{1}{c|}{\textbf{This work}} \\ \hline
		\multicolumn{1}{c|}{\textbf{ $^2S_{1/2}$}} & \multicolumn{1}{c|}{98.77\% }& \multicolumn{1}{c|}{98.86\%}& \multicolumn{1}{c|}{99.09\%}& \multicolumn{1}{c|}{98.86\%}& \multicolumn{1}{c|}{98.83\%}& \multicolumn{1}{c|}{\textbf{98.75(6)\%}} \\  \hline
		\multicolumn{1}{c|}{\textbf{ $^2D_{3/2}$}} & \multicolumn{1}{c|}{0.21\%}& \multicolumn{1}{c|}{0.18\%}& \multicolumn{1}{c|}{0.15\%}& \multicolumn{1}{c|}{0.18\%}& \multicolumn{1}{c|}{0.19\%}& \multicolumn{1}{c|}{\textbf{0.17(1)\%}} \\  \hline
		\multicolumn{1}{c|}{\textbf{ $^2D_{5/2}$}} & \multicolumn{1}{c|}{1.02\%}& \multicolumn{1}{c|}{0.96\%}& \multicolumn{1}{c|}{0.76\%}& \multicolumn{1}{c|}{0.96\%}& \multicolumn{1}{c|}{0.98\%}& \multicolumn{1}{c|}{\textbf{1.08(5)\%}} \\  \hline
		
	\end{tabular}
	\caption{Branching fractions of the decay from the $^2P_{3/2}$ state: Our work and the theoretical (Th.) predictions. In Ref.\,\cite{Porsev:2012} results from a single-electron approach ($a$) and a many-electron approach ($b$) are given. Our values agree roughly with theoretical predictions. }
	\label{tab:BranchingRatio}
\end{table*}

To measure the branching fractions for decay out of the $^2P_{3/2}$ state, we excite the D2-transition with a short pulse of resonant light, followed by fluorescence detection of 100\,$\mu$s duration. During the detection, an ion initially in the $^2S_{1/2}$ or $^2D_{3/2}$ state scatters light, while an ion in the $^2D_{5/2}$ state appears dark. Scanning the pulse width, we obtain the time constant $\tau_{\mathrm{dark}} = 1.04(4)\,\mu$s for the exponential decay of the PMT counts versus pulse width plotted in Fig.\,\ref{fig:Branching}. From this time constant, the lifetime of the $^2P_{3/2}$ state of $\tau_{\mathrm{p32}}=6.15(9)$\,ns\,\cite{Pinnington:1997} and the probability to be in the excited state during the laser-pulse $p_{p32}$, we determine the combined decay probability to both $D$ states as follows:

\begin{equation}\label{key}
p(D\textnormal{-state})=\frac{\tau_{\mathrm{p32}}}{p_{\mathrm{p32}} \times \tau_{\mathrm{dark}}}
\end{equation}

\noindent The saturation parameter $s$ and thus the excited state population probability is determined by a frequency scan over the resonance.
We normalize the power in the 329\,nm beam during the scan by appropriate power settings of the AOM as described above.
We choose a short pulse width of $t_{329} = 500$\,ns to avoid saturation of the decay to the $^2D_{5/2}$ state.  We measure a saturation parameter of $s=11$, the data is shown in the inset of Fig.\,\ref{fig:Branching}. For the branching ratio experiment we tune the laser to resonance and operate the AOM at the frequency of maximum diffraction efficiency which allows us to increase the power in the laser pulse by a factor of 2.2 compared to the scan. Accordingly we calculate a saturation parameter of $s=24(13)$ for the branching ratio experiment, corresponding to a probability of being in the excited state of $p_{\mathrm{p32}} = 0.48(1)$. This scan is repeated frequently during the measurement to monitor intensity and frequency drifts.

The ratio between the decay to the $^2D_{5/2}$ state and $^2D_{3/2}$ state is given by the ratio between the fluorescence for a pulse width $t_{329} = 0$, all population in the $^2S_{1/2}$ state ($n(t_{329}=0) = 0.741$), compared to the fluorescence when all population is pumped to the $D$ manifold ($n(t_{329}=\infty) = 0.104$), as indicated in Fig.\,\ref{fig:Branching}.

We determine branching fractions to the $^2D_{5/2}$ state and $^2D_{3/2}$ state of 1.08(5)\% and 0.17(1)\% respectively which is in agreement with theoretical predictions from Ref.\,\cite{Biemont:1998}.
The errors include the statistical uncertainties in $\tau_{\mathrm{dark}}$, $n(t_{329}=0)$, $n(t_{329}=\infty)$ as well as uncertainties in the excited state population and the lifetime of the $^2P_{3/2}$ state.

The cumulative duration of the excitation pulse ($t_{329} < 12\,\mu$s) and the detection bin ($t_{\mathrm{det}} = 100\,\mu$s) is small compared to the lifetimes of the $^2D_{3/2}$ ($\tau = 52.7$\,ms\,\cite{Yu:2000}) and $^2D_{5/2}$ ($\tau = 7.2$\,ms\,\cite{Taylor:1997}) states. The probability for a decay back to the ground state during that time is less than 0.3\% and thus negligible compared to other experimental errors.

\section{Conclusions}

The isotope shifts in the $^2S_{1/2} \rightarrow\,^2P_{3/2}$ transition in Yb$^+$ and hyperfine splitting (in $^{171}$Yb$^+$) of the $^2P_{3/2}$ state have been determined. Our results on both agree with previous results from Ref.\,\cite{Berends:1992} but are more precise by a factor of 5-9. Both experiments contradict theoretical predictions obtained from single-electron approaches for the hyperfine splitting by a factor 2-3. The results from a many-electron approach\,\cite{Porsev:2012} agree much better with the experimental results. This corroborates their hypothesis that mixing with the energetically close state $4f^{13}5d6s^3\left[3/2\right]^0_{3/2}$ has a strong influence on the $^2P_{3/2}$ state, which can not be described by a single-electron approach. 
We find the branching fractions for decay of the $^2P_{3/2}$ state to the $^2D_{5/2}$ state and $^2D_{3/2}$ states to be 1.08\% and 0.17\%, respectively, in rough agreement with theoretical predictions.

\begin{acknowledgements}
We thank M. Safronova for helpful comments on the manuscript. This work was supported by the European Union via the European Research Council (Starting Grant 337638) and the Netherlands Organization for Scientific Research (Vidi Grant 680-47-538) (R.G.).
\end{acknowledgements}

\section*{References}


%

\end{document}